\documentclass[ twocolumn,  aps,  superscriptaddress,showpacs, pre]{revtex4-1}
\usepackage{graphicx,amsfonts,amsmath,color,amsbsy,amssymb, subfigure}
\usepackage{epstopdf}
\usepackage{float}

\begin{document}

\title{Mass Generation from Embedding Geometry in Surface Nematics}

\author{J.A. Santiago}\email{jsantiago@cua.uam.mx}
\affiliation{Departamento de Matem\'aticas Aplicadas y Sistemas\\ 
Universidad Aut\'onoma Metropolitana Cuajimalpa\\
Vasco de Quiroga 4871, 05348 Cd. de M\'exico, Mexico}
\affiliation{Departamento de Qu\'imica F\'isica, 
Universidad Complutense de Madrid, Ciudad Universitaria s/n, 28040 Madrid, Spain}

\author{F. Monroy}\email{monroy@ucm.es}
\affiliation{Departamento de Qu\'imica F\'isica, 
Universidad Complutense de Madrid, Ciudad Universitaria s/n, 28040 Madrid, Spain}
\affiliation{Institute for Biomedical Research Hospital Doce de Octubre (imas12)\\
Av. Andaluc\'ia s/n, 28041 Madrid, Spain}

\vspace{10pt}

\begin{abstract}
We show that a nematic field constrained to a curved embedded surface develops an emergent geometric mass in its leading isotropic interaction sector. An auxiliary embedding-space closure mediated by the surface spin connection yields a massive scalar mode \(\chi_n\) with mass set by the extrinsic curvature invariant \(m^2=K_{ab}K^{ab}\). This mass arises directly from embedding geometry, promoting the intrinsic massless nematic interaction into a geometry-controlled massive field. The resulting theory identifies Gaussian curvature as a distributed geometric charge and establishes embedding geometry as the regulator of defect interactions on curved nematic membranes.
\end{abstract}

\maketitle

Nematic fields constrained to curved embedded surfaces couple orientational order directly to geometry through tangency and embedding~\cite{Kamien2002,Vitelli2012,Napoli2012}. Beyond intrinsic distortions within the tangent plane, embedding introduces couplings to extrinsic curvature, so that the orientational response depends not only on the surface metric but also on how the surface bends in the ambient space. A natural elastic description is provided by Frank-type theories adapted to tangent nematic fields on embedded surfaces~\cite{Frank1958,Kamien2002,SantiagoMonroyPRE,SantiagoMonroyJPA}. For a director constrained to remain tangent to the surface,
\(
\boldsymbol{\eta}=\eta^a\mathbf e_a
\)
with
\(
\boldsymbol{\eta}\cdot\mathbf n=0,
\)
the projected distortion modes become
\begin{equation}
\nabla\cdot\boldsymbol{\eta}
=
\nabla_a\eta^a,
\qquad
\boldsymbol{\eta}\cdot\nabla\times\boldsymbol{\eta}
=
K_{ab}\eta^a\eta_\perp^{\,b},
\label{eq:Frank_tangent_modes_1}
\end{equation}
and
\begin{equation}
(\boldsymbol{\eta}\cdot\nabla)\boldsymbol{\eta}
=
(\eta^a\nabla_a\eta^b)\mathbf e_b
-
K_{ab}\eta^a\eta^b\,\mathbf n.
\label{eq:Frank_tangent_modes_2}
\end{equation}
Splay remains intrinsic, twist is purely extrinsic, and bend mixes in-surface transport with curvature projection~\cite{Santiago2018PRE,SantiagoMonroyJPA}. Tangency therefore couples orientational elasticity directly to embedding geometry. This raises a central question: is embedding curvature merely a local orientational bias, or can it promote the interaction sector itself into a genuinely massive geometric field?

Focusing on the leading isotropic interaction sector, in which splay, twist, and bend contribute on equal footing~\cite{Frank1958}, the embedded tangent nematic field admits an embedding-space closure whose normal projection yields a massive scalar mode \(\chi_n\) with geometric mass
\[
m^2(\mathbf{x})=K_{ab}K^{ab}=K^2(\mathbf{x})-2K_G(\mathbf{x}),
\]
where \(K_{ab}\) is the second fundamental form of the embedded surface, and \(K\) and \(K_G\) are the mean and Gaussian curvatures~\cite{Kamien2002,docarmo}. This mass is not introduced phenomenologically: it emerges as the geometric zeroth-order term generated by the normal projection of the closure. The same construction identifies topological defects as localized charges and Gaussian curvature as a distributed background charge~\cite{Vitelli2006,Bowick2009,Napoli2012,Mbanga2012}, while extrinsic curvature enters through \(m^2\) to set the interaction scale. Embedding geometry thus promotes the intrinsic massless interaction into a geometry-controlled massive field (Fig.~\ref{fig:Feynman}); derivations are given in the Supplementary Information. At the operator level, embedding fixes a nontrivial spin connection with curvature \( \epsilon^{ab}\nabla_a\Omega_b=K_G \)~\cite{Kamien2002}, which cannot generally be eliminated globally and therefore leaves a geometric imprint on the projected theory. In the leading isotropic interaction sector identified below, this imprint appears as a Helmholtz operator,
\[
-\Delta \;\longrightarrow\; -\Delta + m^2(x),
\]
so that the intrinsic Frank response becomes geometrically screened. In this sense, curvature lifts a massless mode into a massive one in a manner formally analogous to Anderson--Higgs mass generation~\cite{Anderson1963,Higgs1964}, although here the effect is purely geometric and embedding-driven.

\begin{figure}[t]
\centering
\includegraphics[width=\columnwidth]{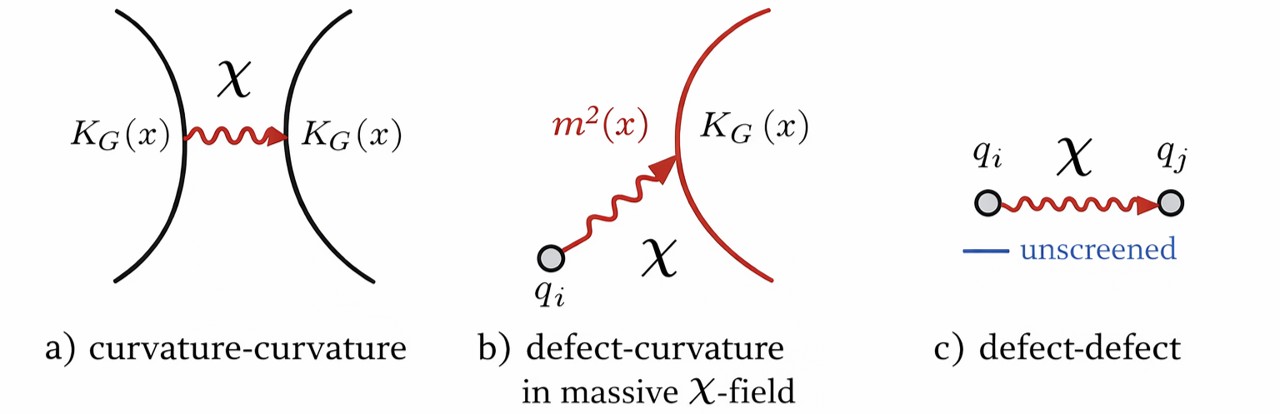}
\caption{
Interaction channels mediated by the emergent massive scalar field \( \chi_n \). (a) Curvature--curvature coupling: nonlocal interaction between regions of Gaussian curvature \(K_G(x)\). (b) Defect--curvature coupling: interaction between a topological charge \(q_i\) and the distributed background, mediated by the curvature-induced mass \(m^2(x)\). (c) Defect--defect coupling: screened elastic interaction between defect charges, controlled by the same geometric mass \(m^2(x)\). In the limit \(m^2=0\), the intrinsic unscreened theory is recovered.
}
\label{fig:Feynman}
\vspace{-10pt}
\end{figure}

Figure~\ref{fig:Feynman} summarizes the resulting physics. The curvature-induced massive mode \(\chi_n\) mediates defect--defect, defect--curvature, and curvature--curvature interactions through a single screened kernel. This mass--charge formulation recasts curvature--nematic interactions and enables a systematic expansion about the intrinsic massless theory, where spin-connection-induced screening is absent~\cite{Kamien2002}. The plane thus provides the genuine massless reference, \(m^2=0\), with long-range intrinsic interactions, whereas curved embedded surfaces acquire a geometry-controlled interaction range, \(m^2\neq0\). We further illustrate the mechanism on the cylinder and sphere, where uniform curvature yields constant mass and explicit screened energetics.\\

\textbf{Geometry-Induced Mass from Embedded Nematic Closure.}
We consider a nematic director \( \boldsymbol{\eta} \) constrained to remain tangent to a two-dimensional surface \( \mathcal{S} \subset \mathbb{R}^3 \). The surface is parametrized by local coordinates \( \xi^a \) through an embedding \( \mathbf{X}(\xi^a) \), with tangent basis \( \mathbf{e}_a=\partial_a\mathbf{X} \) and unit normal \( \mathbf{n} \). In a local orthonormal frame, the director can be written as
\[
\boldsymbol{\eta}=\cos\Theta\,\mathbf{e}_1+\sin\Theta\,\mathbf{e}_2,
\]
where \( \Theta(\xi) \) is the in-surface orientation angle (Supplementary Information, Secs.~I--II). Restricting attention to the leading isotropic interaction sector, in which splay, twist, and bend contribute on equal footing (\(\kappa_1=\kappa_2=\kappa_3\equiv\kappa\)), the embedded tangent nematic field separates into two geometrically distinct contributions (Supplementary Information, Secs.~III--IV): an isotropic covariant-gradient sector and a local curvature-dependent orientational sector. The theory developed here focuses on the first, since it is this sector that generates the massive field \( \chi_n \); the local orientational sector remains part of the full embedded theory, but lies outside the reduced scalar closure derived below. Its energy is
\begin{equation}
    H_{\mathrm{grad}}=\frac{\kappa}{2}\int dA\, g^{ab}(\partial_a\Theta-\Omega_a)(\partial_b\Theta-\Omega_b),
    \label{eq:projected_frank}
\end{equation}
where \(g^{ab}\) is the inverse surface metric and \(\Omega_a\) is the spin connection associated with the rotation of the surface frame~\cite{Napoli2012}. Equation~\eqref{eq:projected_frank} is formally intrinsic, but its closure is not: the tangency constraint ties the covariant orientational current directly to the embedding geometry. Varying Eq.~\eqref{eq:projected_frank} yields the equilibrium condition \(\nabla_a D^a\Theta=0\), with \(D_a\Theta=\partial_a\Theta-\Omega_a\), so the covariant current is divergence-free. This motivates the auxiliary embedding-space representation
\begin{equation}
    \nabla\Theta-\boldsymbol{\Omega}=\nabla\times\boldsymbol{\chi},
    \label{eq:chi_def}
\end{equation}
with \(\boldsymbol{\Omega}=\Omega_a\mathbf{e}^a\). The auxiliary field decomposes as
\[
\boldsymbol{\chi}=\chi^a\mathbf{e}_a+\chi_n\mathbf{n}.
\]
Here the tangential components \( \chi^a \) encode geometric compatibility, while the normal component \( \chi_n \) defines the massive scalar mode governing the leading isotropic interaction sector (Supplementary Information, Sec.~V). Projecting the auxiliary closure onto the surface normal isolates this reduced scalar mode and generates the geometric zeroth-order operator associated with the Gauss invariant \(m^2\equiv K_{ab}K^{ab}\). The ambient curl--Laplacian identity then yields
\begin{equation}
    \left[-\nabla^2+m^2(x)\right]\chi_n(x)=\rho(x),
    \label{eq:chi_field_eq}
\end{equation}
with curvature-induced mass
\begin{equation}
    m^2(x)=K^2(x)-2K_G(x)=K_{ab}K^{ab},
    \label{eq:mass_term}
\end{equation}
where \(K\) and \(K_G\) are the local mean and Gaussian curvatures~\cite{docarmo}. The mass is therefore not introduced phenomenologically, but emerges directly from the normal projection of the ambient vector Laplacian as the extrinsic quadratic invariant of the embedding. The corresponding source term is
\begin{equation}
    \rho(x)=2\pi\sum_i q_i\,\delta_S(x,x_i)-K_G(x),
    \label{eq:rho_def}
\end{equation}
where \(\delta_S(x,x_i)\) denotes the surface Dirac distribution, so that the first term represents localized topological charges \(q_i\), while the second is a distributed Gaussian-curvature background charge~\cite{Vitelli2006,Bowick2009,Napoli2012,Mbanga2012}. 

Equation~\eqref{eq:chi_field_eq} is therefore the central field equation of the present theory: the curvature-induced mode \( \chi_n \) obeys an inhomogeneous Helmholtz equation even though the original orientational energy [Eq.~\eqref{eq:Frank_general}] contains no explicit mass term. Embedding geometry thus lifts the massless character of the covariant-gradient sector and endows the interaction kernel with a finite geometric scale \( \ell_m \sim m^{-1} \). The scalar closure is controlled in the adiabatic regime of weak curvature gradients, \( |\nabla K| \ll |K|^2 \), where the normal mode dominates the long-wavelength massive response (Supplementary Information, Sec.~VI).\\

\begin{figure}[b!]
\centering 
\includegraphics[scale=0.37]{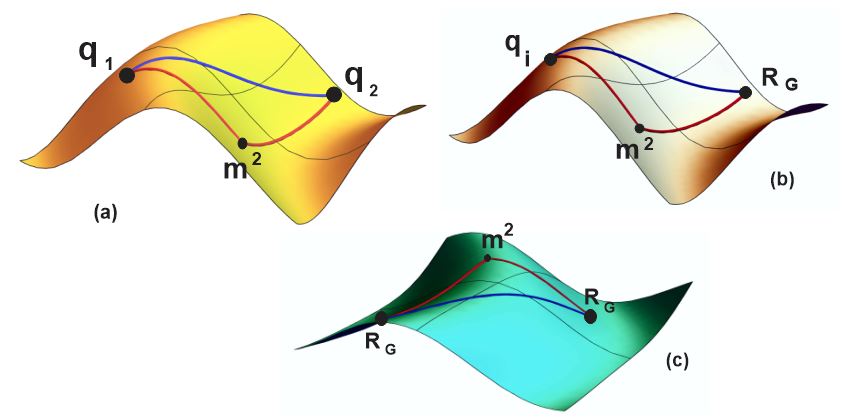}
\caption{
Three interaction channels generated by the on-shell energy of the massive field \(\chi_n\) on a curved surface. 
(a) Defect--defect interaction \(F_{dd}\): two topological charges \(q_1\) and \(q_2\) couple through the Green kernel \(G(x,x')\). 
(b) Defect--curvature interaction \(F_{dK}\): a defect charge \(q_i\) couples to the Gaussian-curvature background \(K_G(x)\). 
(c) Curvature--curvature interaction \(F_{KK}\): nonlocal coupling between curvature regions mediated by the same kernel. 
In all cases, the interaction strength and scale are controlled by the geometric mass \(m^2(x)\).
}
\label{fig:QQQ}
\vspace{-2mm}
\end{figure}

\textbf{Defect Energetics: Massive Interaction Channels.}
Topological defects on curved surfaces act as localized singularities of the director angle \( \Theta(x) \), with
\begin{equation}
    \nabla \times \nabla \Theta(x) = 2\pi \sum_i q_i\, \delta_S(x,x_i),
\end{equation}
where \( q_i \) is the topological charge of a defect at \( x_i \) (Supplementary Information, Secs.~V and VII). In the present formulation, these sources couple through the massive field \( \chi_n \), which obeys Eq.~\eqref{eq:chi_field_eq} with total source
\[
\rho(x) = 2\pi \sum_i q_i\, \delta_S(x,x_i) - K_G(x),
\]
and geometric mass \( m^2(x)=K^2(x)-2K_G(x)=K_{ab}K^{ab}(x) \). The corresponding Green function \( G(x,x') \) satisfies
\begin{equation}
    \left[-\nabla^2 + m^2(x)\right] G(x, x') = \delta_S(x,x'),
\end{equation}
so that
\begin{equation}
    \chi_n(x) = \int dA'\, G(x, x')\, \rho(x').
\end{equation}
Substituting this solution back into the embedded interaction energy, the on-shell contribution becomes
\begin{equation}
    F_{\chi} = \frac{\kappa}{2}\int dA\, \chi_n(x)\, \rho(x),
\end{equation}
which reorganizes the interaction sector into three geometrically distinct channels,
\begin{align}
    F_{\chi} &= F_{dd} + F_{dK} + F_{KK}, \label{eq:F_total} \\
    F_{dd} &= 2\pi^{2}\kappa \sum_{i,j} q_i q_j\, G(x_i, x_j), \\
    F_{dK} &= -2\pi\kappa \sum_i q_i \int dA\, G(x_i, x)\, K_G(x), \\
    F_{KK} &= \frac{\kappa}{2}\int dA\, dA'\, K_G(x)\, G(x,x')\, K_G(x').
\end{align}
The same massive kernel \(G(m;x)\) therefore governs defect--defect (\textit{dd}), defect--curvature (\textit{dK}), and curvature--curvature (\textit{KK}) interactions (Fig.~\ref{fig:QQQ}). All three are nonlocal and modulated by the local mass \(m^2(x)\), which sets their geometry-dependent interaction scale. In the weak-mass regime, \(L\ll \ell_m\equiv m^{-1}\), or equivalently \(m^2L^2\ll1\), the theory reduces perturbatively to the intrinsic massless limit. The Green kernel then admits the expansion (Supplementary Information, Sec.~VIII)
\begin{equation}
    G \simeq G_0 - G_0 m^2 G_0 + G_0 m^2 G_0 m^2 G_0 + \cdots,
\end{equation}
around the massless propagator \(G_0\), yielding
\begin{equation}
    F_{\chi} = F^{(0)} + \delta F^{(1)} + \delta F^{(2)} + \cdots,
\end{equation}
where \(F^{(0)}\) is the intrinsic massless contribution and higher-order terms encode geometric mass corrections of order \(L^2/\ell_m^2\)~\cite{Kamien2002}. As shown in the Supplementary Information (Sec.~IX), the leading correction is negative definite, expressing suppression of long-range interactions by the curvature-induced mass. Figure~\ref{fig:QQQ} summarizes the physical content of this decomposition. The curvature-induced field \(\chi_n\) therefore provides a unified description of defect energetics and curvature response, with the local mass \(m^2(x)\) controlling all interaction channels. The theory reduces continuously to the intrinsic massless limit in the flat-membrane case, where \(K^2=K_G=0\) and \(m^2=0\), recovering long-range unscreened interactions.\\

\textbf{Analytical Solutions: Plane, Cylinder, and Sphere.}
The flat membrane provides the genuine massless reference, \(m^2=0\), for which interactions remain long-ranged and are governed by the intrinsic Laplace--Beltrami operator. To isolate the effect of geometry-induced mass, we next consider constant-curvature surfaces, where \(m^2\) is uniform and the field equation for \(\chi_n\) admits analytic solution (Supplementary Information, Sec.~X).

\textbf{Cylinder.}
The cylinder provides the sharpest benchmark for geometric mass generation. Although its Gaussian curvature vanishes, \(K_G=0\), its extrinsic curvature does not: \(K_1=1/R\), \(K_2=0\), so
\begin{equation}
    m_{\mathrm{cyl}}^2=K_{ab}K^{ab}=\frac{1}{R^2}.
\end{equation}
The field \(\chi_n\) therefore obeys a Helmholtz equation with uniform mass, whose exact solution is given in Supplementary Information (Sec.~X.A). The resulting propagator acquires an exponential scale, and at large axial separation is dominated by
\begin{equation}
    G_{\mathrm{cyl}}(s,\phi;s',\phi') \sim \frac{1}{4\pi}e^{-|s-s'|/R},
\end{equation}
so that the interaction length is \(\ell_m=m^{-1}=R\), i.e. Yukawa-like~\cite{Yukawa1935}. The cylinder isolates the central mechanism of the theory: even in the absence of Gaussian curvature, extrinsic curvature alone generates a finite mass and converts the intrinsic long-range interaction into a finite-scale one.

\textbf{Sphere.}
On the sphere, the massive response is spectral rather than real-space. For a sphere of radius \(R\), the principal curvatures are \(K_1=K_2=1/R\), so that
\begin{equation}
    K_G=\frac{1}{R^2},
    \qquad
    m_{\mathrm{sph}}^2=K_{ab}K^{ab}=\frac{2}{R^2}.
\end{equation}
The reduced field equation becomes
\begin{equation}
    \left[-\nabla^2+m_{\mathrm{sph}}^2\right]\chi_n(\Omega)=\rho(\Omega),
\end{equation}
with \(\Omega=(\theta,\varphi)\) and
\begin{equation}
    \rho(\Omega)=2\pi\sum_i q_i\,\delta_S(\Omega,\Omega_i)-\frac{1}{R^2}.
\end{equation}
Its exact solution follows from spherical-harmonic expansion (Supplementary Information, Sec.~X.B). Since
\begin{equation}
    -\nabla^2 Y_{\ell m}=\frac{\ell(\ell+1)}{R^2}Y_{\ell m},
\end{equation}
the Green function is
\begin{equation}
    G_{\mathrm{sph}}(\Omega,\Omega')
    =
    \sum_{\ell,m}
    \frac{Y_{\ell m}(\Omega)Y_{\ell m}^*(\Omega')}
    {\ell(\ell+1)/R^2+m_{\mathrm{sph}}^2},
\end{equation}
or equivalently, in terms of the geodesic distance \(\gamma\),
\begin{equation}
    G_{\mathrm{sph}}(\gamma)
    =
    \frac{1}{4\pi}
    \sum_{\ell=0}^{\infty}
    \frac{2\ell+1}{\ell(\ell+1)/R^2+m_{\mathrm{sph}}^2}
    P_\ell(\cos\gamma),
\end{equation}
with \(m_{\mathrm{sph}}^2=2/R^2\)~\cite{Szmytkowski2006,Grigoryan2009}. Unlike the cylinder, the sphere exhibits no real-space Yukawa tail because the surface is compact. The curvature-generated mass instead reorganizes the modal structure: it suppresses the low-\(\ell\) response, removes the Laplacian zero mode, and opens a finite spectral gap in the interaction kernel.\\

\textbf{Conclusion}. We have shown that constraining a nematic director to a curved embedded surface generates, within the leading isotropic interaction sector of an embedded tangent nematic field, a geometric mass
\(
m^2 = K_{ab}K^{ab} = K^2 - 2K_G
\)
for the emergent scalar mode \(\chi_n\). This massive field arises as the normal projection of the covariant-current closure of the embedded nematic sector, encoding how the spin connection converts embedding geometry into an interaction field coupled to the director. Only the flat membrane (\(K_{ab}=0\), hence \(m^2=0\)) recovers the genuine massless Coulomb-like isotropic structure. Curved embeddings, by contrast, lift that intrinsic response into a geometry-controlled massive interaction. The theory thus acquires a unified mass--charge structure: Gaussian curvature acts as a distributed background charge, defects act as localized charges, and \(\chi_n\) mediates all couplings through a single massive kernel. The cylinder and sphere realize this mechanism in two complementary ways: exponential real-space decay on the cylinder, and spectral gap opening on the sphere. In this sense, mass generation in the covariant-gradient sector is formally analogous to Anderson--Higgs mechanisms: a massless interaction mode acquires a finite scale through coupling to a background field, although here that field is curvature itself rather than a condensate. The resulting mass \(m\) endows the interaction sector with a geometric scale \(\ell_m\sim m^{-1}\), converting a scale-free long-range response into one that is finite-scale on open surfaces and spectrally gapped on closed ones. The mass originates from the normal projection of the covariant orientational current, which couples the spin connection directly to embedding curvature.\\

\textbf{Outlook}. The geometric mass uncovered here has direct physical consequences. Embedding curvature renormalizes the scale of topological interactions, driving a crossover from massless long-range coupling on flat membranes to finite-scale interactions on curved surfaces; on the cylinder, this persists even at \(K_G=0\), showing that extrinsic curvature alone generates mass. On the sphere, the same mechanism becomes spectral, suppressing low modes and opening a gap in the interaction kernel. More generally, the scalar sector already defines a curvature-controlled field theory in which defects and Gaussian curvature couple through a common massive propagator. Extending the closure to the in-plane sector should generate anisotropic kernels and provide a route toward fully vectorial descriptions of curved nematic interactions.\\

This research was supported by SECIHTI-Mexico (Grant CBF-2025-I-4418), AEI-Spain (Grants TED2021-132296B-C52 and CPP2024-011880), and Fundación BBVA (Programa Fundamentos 2025).
 
\bibliography{References}

\end{document}